# Importance of Dispersion and Relativistic Effects for ORR Overpotential Calculation on Pt(111) surface

Akhil S. Nair,[†] Biswarup Pathak[†, #, *]

[†]Discipline of Chemistry, Indian Institute of Technology Indore, Simrol, Indore 453552, India

[#]Discipline of Metallurgy Engineering and Materials Science, Indian Institute of Technology Indore, Simrol, Indore 453552, India

Email: biswarup@iiti.ac.in

**Abstract:** Density functional theory (DFT) has been used as an important tool for studying activity of oxygen reduction reaction (ORR) catalysts. The dispersion effects, which are not encountered in many of the previous DFT studies for periodic Pt(111), are scrutinized for their role in predicting ORR activity on Pt (111) surface. Spin orbit coupling is employed to account for relativistic effects expected for heavy metal platinum, which has not been addressed in any of the previous studies on Pt(111). Adsorption behavior of intermediates and free energy changes of elementary reactions of ORR are analyzed with commonly used dispersion methods. A cumulative enhancement of ORR energetics and a maximum of 25% improvement in theoretical limiting potential are observed. The study illustrates the importance of consideration of these effects for better prediction of electrocatalytic activity for platinum based catalysts.

# 1. <u>Introduction</u>

Electrocatalysis has been appraised as an area of tremendous importance owing to its supreme role in empowering the development of renewable energy related materials to meet the proliferating energy demand.[1,2] Very recent decades have witnessed profound advancement in this area with the introduction of fuel cells, batteries, hydrogen storage materials and photocatalysts to resolve the burgeoning energy crisis by successfully utilizing the results obtained from extensive studies on electrocatalytic reactions.[3-5] Oxygen reduction reaction (ORR) is one of the well-known as well as widely studied electrocatalytic reactions occurring in the cathode of proton exchange membrane (PEM) fuel cell. It is the sluggish nature of ORR which renders challenges in attaining the expected efficiency of the fuel cell in spite of its desirable properties like high thermodynamic efficiency, energy density and zero emission.[6-8] Pt based catalysts retain the dominancy in fuel cell catalysis regime irrespective of development of several non-Pt catalysts because of the vulnerable nature of later in fuel cell working conditions.[9] Therefore scrutinizing ORR activity of Pt based catalysts is a core objective of fuel cell research. Ample attempts to unravel the complex mechanism and improve the energetics of ORR have been carried out so forth among which computational studies, especially density functional theory (DFT) has contributed to a significant extent. For example, the first theoretical explanation for the ORR overpotential generation was given by Norskov et al.[10] grounded on the binding energy descriptor-based approach along with a screening of activity of various catalysts as a volcano plot. An impactful experimental validation of this study was done by Stamkeovic et al.[11] by synthesizing alloy based catalysts with lesser binding energy for ORR intermediates in 0.1M HClO$_4$ at room temperature with 1600 rpm and reported that single crystal surface of Pt$_3$Ni(111) shows 10 times higher ORR activity than Pt(111) surface and 90 times higher than Pt/C catalysts. This has set off numerous experiments for developing efficient catalysts by alloying Pt with carefully selected other metals. Similarly, Kattel and Wang in their DFT study,[12] observed a lowering of binding energy of intermediates and activation of elementary steps of ORR on a strained Pt (111) surface as compared to the unstrained surface. This and similar theoretical studies have paved way to the development of core-shell, heteroatom doped and nanocluster based catalysts which have been found to enhance the ORR energetics stemming from possession of strained surfaces.[13-15] Moreover, several theoretical studies have been helpful for underpinning the complex mechanism and analyzing the kinetics along different pathways of ORR.[16,17] Thus, considerable extent of experimental validations and legitimate explanations of catalytic activity suggest the requisite importance of theoretical studies in exploring the new dimensions of fuel cell catalysis.

The state of art DFT studies have been successful in predicting many ORR related parameters such as adsorption energy, activation barrier, structural dependent activity and so on. Although a complete description of ORR requires explicit modelling of electrode-electrolyte interface, gas phase theoretical studies have been fairly useful in predicting the energetics of ORR and being used extensively to explain the activity of bulk as well as nanocluster catalysts. One of the important drawbacks with these DFT methods is the inadequacy of exchange-correlation (XC) functionals to account for long range dispersion interactions which are vital in providing comprehensive adsorption scenario.[18] Different models and methods have been proposed and widely used for treating dispersion interactions in surface catalysis studies. Grimmie et al. proposed the DFT-D2[19] and DFT-D3[20] methods, which involve addition of empirical correction using interatomic potentials for a better description of long range interactions. Later nonlocal van der Waals density functional (vdw-DF) methods were introduced by Dion et al.[21] which have reported to work well for layered systems such as graphite, metal organic frameworks whereas DFT-D methods for adsorption on transition metals .[22] Although no exclusive physisorption of ORR intermediates are occurring on Pt, inclusion of dispersion interactions is necessary for acquiring accurate description of surface adsorbate energetics. Another noticeable factor in heavy metal mediated catalysis is the relativistic effects which may alter the electronic state of the metal atoms and thereby can be expected to modify catalytic properties. Although often computational calculations employ the use of relativistic pseudo potentials to simulate the core electrons, consideration of nonscalar relativistic effects such as spin orbit coupling (SOC) have been identified to bring out significant changes in the structural as well as activity properties of heavy elements.[23,24] The heavy metals such as gold, lead and lanthanides have been extensively studied for their relativistic effects while similar extent of studies have not been reported in platinum.[25-27] SOC has reported to play decisive role in CO adsorption on Ni and Pt surfaces.[28] Majority of previously reported studies were focused on structural changes induced by SOC on small platinum clusters ($Pt_n$, n=2-5) and the effects in catalytic activity are less explored.[29,30]

Our current study delves into the effect of dispersion interactions and relativistic considerations in determining the energetics of ORR on a periodic Pt (111) surface. We have chosen the periodic Pt (111) system as it is the most extensively studied system for ORR and hence a comparison with previous experimental as well as theoretical results are possible. Theoretical limiting potential is selected in the work as the most important benchmark parameter as it correlates the catalytic activity via free energies of elementary steps of ORR and it is an important parameter experimentalists looking for. We systemically analyze the adsorption of important ORR intermediates as well as the energetics of electrochemical steps under applied potential with different dispersion correction methods with different levels of theory. The effect of spin orbit coupling in the reaction energetics is investigated and compared the results with non-SOC data. The relevance of these effects in theoretical calculations is scrutinized and implications for future studies are discussed.

## 2. **Computational Methodology**

The DFT calculations are carried out using VASP (Vienna Ab Initio Software Pacakge) [31] with projector augmented wave (PAW) method[32] under periodic boundary conditions. Generalized gradient approximations of Perdew–Burke–Ernzerhof (GGA-PBE)[33] and revised PBE(GGA-RPBE)[34] are used for describing the exchange correlation interactions. Semi empirical dispersion corrections with DFT-D2 and DFT-D3 methods in both PBE and RPBE level and nonlocal vdw-DF methods implemented in the VASP code are adapted for comparative studies. In the DFT-D2 method[19], the dispersion included total energy is given by

$$E_{DFT-D2} = E_{KS-DFT} + E_{D2}^{disp} \qquad (1)$$

where, $E_{KS-DFT}$ is the energy obtained by solving Kohn-Sham equations at a chosen exchange correlational functional (XC) and $E_{disp}$ is an empirical correction given by,

$$E_{D2}^{disp} = -s_6 \sum_{i=1}^{N_{at}-1} \sum_{j=i+1}^{N_{at}} \frac{C_6^{ij}}{R_{ij}^6} f_{dmp}(R_{ij}) \qquad (2)$$

where, $s_6$ is a scaling factor depending on the XC functional, $N_{at}$ is the number of atoms, $C_6^{ij}$ is the coefficient of dispersion for a pair of atoms i, j and $R_{ij}$ is the distance between atoms. $f_{dmp}(R_{ij})$ stands for a damping function to eliminate singularities at short values of $R_{ij}$. The DFT-D3 method[20] involves a modification in the DFT-D2 method by adapting dispersion correction as a sum of two or three body energies which is given by,

$$E_{DFT-D3} = E_{KS-DFT} - E_{D3}^{disp}, \text{ where} \quad (3)$$

$$E_{D3}^{disp} = E_{D2}^{disp} + E^3,$$

with the three-body energy term is given by,

$$E^3 = \sum_{ABC} f_{d,(3)}(r_{ABC}) E^{ABC} \quad (4)$$

where, the sum extends over all triples of atoms ABC in the system. The damping function is described by geometric average radii $r_{ABC}$ of the system. The three body energy term is more pronounced in systems with large number of atoms. The computational cost does not rise to higher orders even after the inclusion of D3 correction as it is majorly contributed by the time required to solve the Kohn-Sham equations. The optimized PBE (optPBE) and optimized Becke88 (optB88) are examples for van der waals-Density Functionals (vdw-DF)[35] on contrast to semi empirical corrections as in DFT-D methods, include a nonlocal correction term added to the exchange correlational functional which can be represented as,

$$E_{xc}[n] = E_x^{revPBE}[n] + E_c^{LDA}[n] + E_c^{nl}[n] \quad (5)$$

where the first two terms in the right hand side stand for revPBE exchange and LDA correlation respectively and the last term represents the nonlocal correction which accounts for the dispersion corrections. These methods require a relatively higher computational cost as compared to DFT-D methods.

The spin orbit coupling hamiitonian $\hat{H}_{SO}$ in PAW method[36] can be represented as,

$$\widetilde{H}_{SO} = \sum_{ij} |p_i\rangle\langle\Phi_i|H_{SO}|\Phi_j\rangle\langle p_j| \qquad (6)$$

where the Hso, given in zeroth order regular approximatin (ZORA) as ,

$$\widetilde{H}_{SO}^{\alpha,\beta} = \frac{h^2}{(2m_e c)2} \frac{K(r)}{r} \frac{dV(r)}{dr} \sigma^{\alpha,\beta}.L \qquad (7)$$

Here $L$ represents the angular momentum , σ corresponds to spin matrices, $V(r)$ represents spherical part of effective all-electron potential in the PAW sphere, and

$$K(r) = \left(1 - \frac{V(r)}{2m_e c^2}\right)^{-2} \qquad (8)$$

The action of the SOC operator on the Pseudo orbitals can be calculated as,

$$|\tilde{\psi}_n^\alpha\rangle = \sum \widetilde{H}_{SO}^{\alpha\beta} |\tilde{\psi}_n^\beta\rangle \qquad (9)$$

where $\alpha$ and $\beta$ label the spin-up and spin-down components the spinor wave functions.

Plane waves with an energy cutoff of 470 eV are employed for expanding wavefunctions. Conjugate gradient algorithm[37] is used for performing ionic relaxations with a tolerance of $10^{-4}$ eV for minimum energy and 0.02 eV Å$^{-1}$ for Hellmann-Feynman forces on atoms. A 3x3 supercell of periodic Pt (111) surface is constructed using a slab model consisting of 4 layers. The top two layers are relaxed and bottom two layers are frozen during geometry optimization. The lattice constant of the periodic Pt (111) surface is fixed same as the experimental lattice constant of 2.77 Å as we are comparing our results with experimental studies. A vacuum of 15 Å is inserted to avoid interaction between periodic images. Brillioun zone of surface is sampled with 5x5x1 Monkhrost Pack k-point grid for relaxation.[38] The k-point mesh, plane wave energy cutoff and number of atomic layers are tested for convergence. Spin polarized calculations are carried out for studies involving molecular oxygen. The adsorption calculations are performed at 1/9 ML coverage on the surface and keeping sufficient distance between nearby adsorbate atoms.

## 3. Results and Discussions

### 3.1 Dispersion Effects

The adsorption configurations of $O_2^*$, $O^*$, $OH^*$ and $OOH^*$ (* indicates adsorbed species) on the periodic Pt (111) surface are shown in Fig. 1. All the possible sites of occupancy for these species are investigated and the most stable configurations are considered. The favorable sites of adsorption are found to be same with DFT-D methods at PBE and RPBE level and also with vdw-DF methods which are observed to be di-sigma bridge, 3-fold fcc, top, and top sites for $O_2^*$, $O^*$, $OH^*$ and $OOH^*$ respectively. The binding energy of $O^*$, $OH^*$ and $OOH^*$ are calculated with respect to $H_2O$ (l) and $H_2$ (g) using the following equations[10],

$$Pt^* + H_2O(l) \rightarrow Pt - O^* + H_2(g) \quad (10)$$

$$\Delta E_{O*} = E_{Pt-O*} - E_{Pt*} - (E_{H_2O} - E_{H_2})$$

$$Pt^* + H_2O(l) \rightarrow Pt - OH^* + 1/2 H_2(g) \quad (11)$$

$$\Delta E_{OH*} = E_{Pt-OH*} - E_{Pt*} - (E_{H_2O} - 1/2 E_{H_2})$$

$$Pt^* + 2H_2O(l) \rightarrow Pt - OOH^* + 3/2 H_2(g) \quad (12)$$

$$\Delta E_{OOH*} = E_{Pt-OOH*} - E_{Pt*} - (2E_{H_2O} - 3/2 E_{H_2})$$

where $E_{Pt-x*}$ corresponds to energy of adsorbed species and $E_{H_2O}$ and $E_{H_2}$ corresponds of energy of gase phase $H_2O$ and $H_2$ respectively.

**Fig. 1:** Top and side views of adsorption configurations of ORR intermediates, (a) $O_2^*$, (b) $O^*$, (c) $OH^*$, (d) $OOH^*$

The free energies of these intermediates are obtained by incorporating zero point energy and entropy to the binding energy which result in corrections by 0.05 eV, 0.35 eV and 0.40 eV for $O^*$, $OH^*$ and $OOH^*$ species respectively as reported in previous studies.[39] The vibrational frequency calculations have shown negligible difference in zero point energy for all the DFT methods considered and hence the correction factors are kept same for all calculations. The calculated binding energy values are given in Table S1 and a comparison with different DFT methods is represented in Fig. 2.

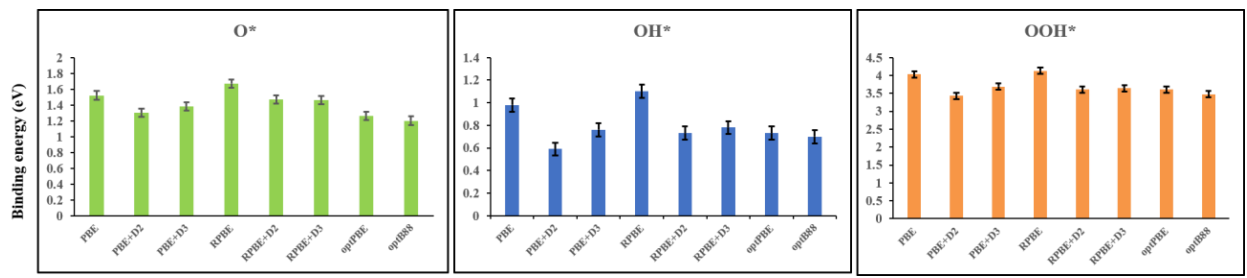

**Fig. 2:** Binding energy of ORR intermediates calculated with different DFT methods.

From Fig. 2, it can be understood that the binding energy calculated as reaction energies of equations 10-12, considerably differ among different DFT methods. According to these reactions, the adsorbed species are formed from $H_2O$ where the lowest endothermicity is for $OH^*$ as it involves the breaking of only one O-H bond. The strongest adsorption of $O^*$ is observed for optB88 vdw functional whereas it is PBE+D2 for $OH^*$ and $OOH^*$ which implies a species dependent adsorption behavior of different DFT methods considered. It is also noticeable that semi- empirical dispersion correction methods D2 and D3 show considerable difference from non-dispersion PBE and RPBE methods. Nevertheless, the trend observed with different dispersion methods is same for O*, OH* and OOH* because of the similar nature of bonding to the Pt (111) surface through O atom consistent with ORR study on large number of metal and metal oxide surfaces.[40-42]

The free energies of adsorption of intermediates obtained by the inclusion of zero point energy and entropy correction have been well known for showing interdependence through scaling relations. The established scaling relations are $\Delta G_{OH}=\Delta G_{OOH} + 3.2 \pm 0.2$ eV for OH* vs OOH* and $\Delta G_{OH}=\Delta G_{O} + 0.2 \pm 0.2$ eV for OH* vs O*.[40,41] Although the scaling relations cannot be derived for a single system but ORR study over a number of distinct systems, the recent reports suggest that it is valid to plot scaling relation between intermediates of catalytic reaction over different methods to obtain the extent of correlation of these methods to the scaling behavior of intermediates. For example, Christensen et al. have reported a systematic error of 0.2 eV in determining the OOH* adsorption energy and changes the scaling offset by 0.1 eV via a study considering different DFT functionals but excluding the often used semi empirical dispersion methods.[42] Similarly, in an oxygen evolution reaction (OER) study with different functionals, Briquet et al. have reproduced similar scaling behavior as that reported in literature.[43] Following these studies, we have checked the plausibility of a scaling behavior between free energy of adsorption of ORR intermediates with the DFT methods considered in the study represented in Fig. 3.

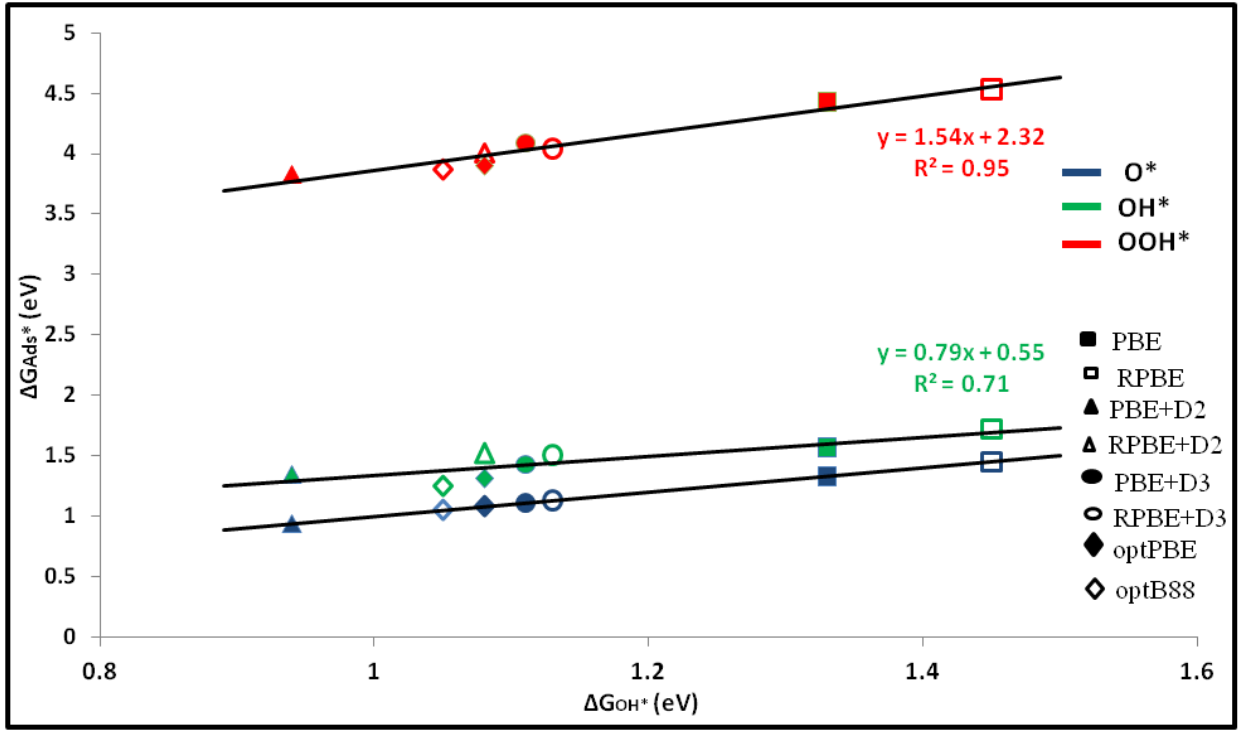

**Fig. 3:** Adsorption free energy relation between $\Delta G_{OH^*}$ vs $\Delta G_{O^*}$ and $\Delta G_{OH^*}$ vs $\Delta G_{OOH^*}$.

From the figure, a reasonable scaling between adsorption free energies of $\Delta G_{O^*}$ and $\Delta G_{OOH^*}$ with $\Delta G_{OH^*}$ is observed. A comparatively better $R^2$ value suggest that the scaling relation is well obeyed between OOH* and OH* as compared to O* and OH*. It is noticeable that the vdw-DF methods optPBE and optB88 show higher deviation from the linear relation as compared to other dispersion correction methods considered which arises due to the strong binding of intermediates calculated at these levels. The sound difference in scaling between O* vs OH* and OOH* vs OH* suggests that the free energy of ORR elementary steps mediated by each pair of reactants would differ significantly over this range of methods. To verify this, we have calculated the free energy of ORR elementary steps in accordance with the following equations;

$$O_{2(g)} + (H^+ + e^-) + * \rightarrow OOH^* \qquad (13)$$

$$OOH^* + (H^+ + e^-) \rightarrow O^* + H_2O \tag{14}$$

$$O^* + (H^+ + e^-) \rightarrow OH^* \tag{15}$$

$$OH^* + (H^+ + e^-) \rightarrow H_2O + * \tag{16}$$

The free energies are calculated as,

$$\Delta G_4 = (\Delta G_{OOH^*} - 0.30) - 4.92 \tag{17}$$

$$\Delta G_5 = (\Delta G_{O^*} - 0.30) - \Delta G_{OOH^*} \tag{18}$$

$$\Delta G_6 = (\Delta G_{OH^*} - 0.30) - \Delta G_{O^*} \tag{19}$$

$$\Delta G_7 = - (\Delta G_{OH^*} - 0.30) \tag{20}$$

where 4.92 eV is the Gibbs free energy of formation of water from $H_2$ and $O_2$ and 0.30 eV is solvent correction corresponding to enthalpy difference between liquid and gaseous states of water. The corresponding free energy diagrams are represented in Fig. S1 in supporting information.

## 3.2 Relativistic Effects

The nonscalar relativistic effects are accounted by the inclusion of spin orbit coupling in calculations. We have optimized all the structures with SOC at different DFT methods considered except vdw-DF methods as they are not properly defined for spin based calculations. By carefully analyzing the binding energy values, it is observed that that the spin orbit coupling does not vary the binding energy values of ORR intermediates to a large extent (see Table S1). However, a slight change in binding energy values for intermediates which are accounted in every elementary step imply that there could be a non-negligible difference in free energy change of ORR elementary steps and hence in overpotential. The changes in binding energy have observed to be of similar extent at PBE, RPBE, DFT-D methods which rules out the possibility of origin of functional dependent errors. Therewith we have calculated the free energy of ORR steps with SOC following equations 13-16, the corresponding free energy diagrams are given in Fig. S2 and a comparison with non-SOC results is depicted in Table 1.

**Table 1:** Free energies of four elementary ORR steps and corresponding rate determining steps.

| DFT method | $\Delta G_{13}$(eV) | | $\Delta G_{14}$(eV) | | $\Delta G_{15}$(eV) | | $\Delta G_{16}$(eV) | | Rate Determining Step | |
|---|---|---|---|---|---|---|---|---|---|---|
| | Non SOC | SOC | Non SOC | SOC | Non SOC | SOC | Non SOC | SOC | Non SOC | SOC |
| PBE | 0.79 | 0.79 | 2.56 | 2.53 | 0.54 | 0.59 | 1.02 | 1.0 | 15 | 15 |
| PBE+D2 | 1.38 | 1.36 | 2.18 | 2.19 | 0.71 | 0.75 | 0.64 | 0.62 | 16 | 16 |
| PBE+D3 | 1.13 | 1.13 | 2.36 | 2.34 | 0.62 | 0.66 | 0.81 | 0.79 | 15 | 15 |
| RPBE | 0.69 | 0.67 | 2.51 | 2.51 | 0.57 | 0.62 | 1.15 | 1.13 | 15 | 15 |
| RPBE+D2 | 1.21 | 1.21 | 2.19 | 2.18 | 0.74 | 0.78 | 0.78 | 0.76 | 15 | 16 |
| RPBE+D3 | 1.18 | 1.18 | 2.23 | 2.21 | 0.68 | 0.74 | 0.83 | 0.79 | 15 | 15 |

From the table, we can see that the free energies calculated for each ORR elemental step differs substantially over DFT methods considered. This is further reflected in the theoretical limiting potential at which all the reaction steps are downhill in energy. For an ideal catalyst, $\Delta G_{13}=\Delta G_{14}=\Delta G_{15}=\Delta G_{16}=$ 1.23 V i.e, all the intermediate steps should have equal free energy change of 1.23 eV.[44,45] But this condition is not met with owing to the fundamental scaling in adsorption energetics of ORR intermediates which leads to an overpotential generation limiting the ORR activity. Now we have observed that for almost all the DFT methods, the rate/potential determining step is OH* formation for which there is always an increment of 0.04-0.06 eV when SOC is included. This can be directly correlated with the theoretical limiting potential which is defined as $U_L= \min(\Delta G_{13},\Delta G_{14},\Delta G_{15},\Delta G_{16})$ at which all the steps are energetically favourable and hence, closer the minimum $\Delta G$ value to 1.23 eV, higher the ORR activity. Hence an increment in $\Delta G_3$ improves the ORR activity by reducing the overpotential by almost 0.05 V. The theoretical limiting potential obtained from different DFT methods with and without SOC consideration are represented in Fig. 4. The maximum improvement in limiting potential by including SOC observed for RPBE-D3 and that is 0.06 V whereas the highest value for limiting potential is observed for RPBE-D2 method with inclusion of SOC and which turns to be 0.76 V. This is the only step which has been observed to have a change in rate determining step when SOC is considered. This change in RDS can be attributed to the comparable free energies observed for OH* formation as well as OH* removal which brings out a maximum limiting potential of 0.76 V. This value closely resembles the limiting potential of periodic Pt (111) surface reported by Norskov et al. (0.78 V) with explicit solvent consideration through bilayer water model.[10] Here it is noticeable that for PBE+D2 level, there occurs a slight decrease in limiting potential arising from deviation of RDS from OH* formation to OH* removal to form water which could be due to its lowest free energy values for H2O formation (RDS) which could not be altered by the improvement in energetics for OH* formation. The improvement in limiting potential observed for PBE and RPBE levels are 0.12 and 0.19 V respectively. These values contribute by 18% and 25% of the observed total ORR limiting potentials of 0.66 and 0.76 V respectively which establishes the importance of dispersion and relativistic effects in ORR catalytic activity calculations. In addition, these values are close to the solvent correction value of 0.30 V employed in free energy calculations which reveals that a correction in terms of dispersion and relativistic effects would be requisite for an improved DFT prediction of platinum based ORR activity which can be verified by further studies involving more dispersion methods as well as platinum based systems.

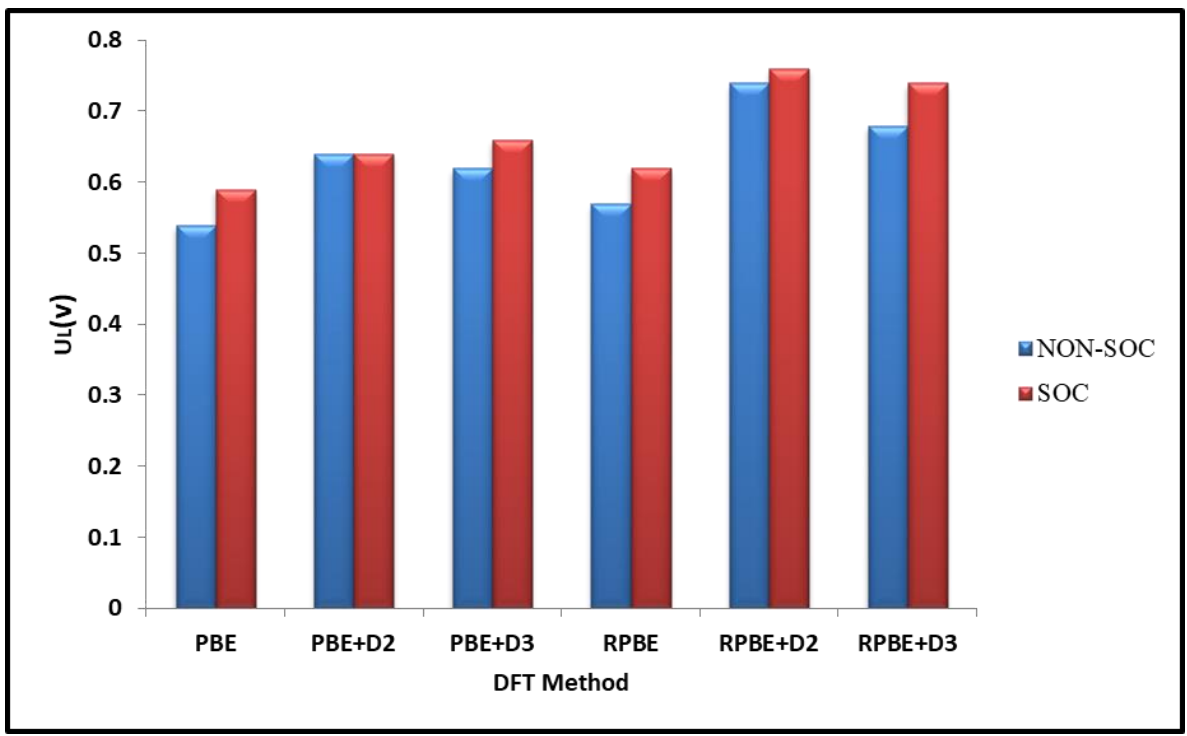

**Fig. 4:** Variation of theoretical limiting potential ($U_L$) across different dispersion methods with and without inclusion of spin orbit coupling (SOC).

## 4. Conclusion

It is evident from this study that the inclusion of dispersion correction is critical in determining the catalytic activity of periodic Pt (111) surface for ORR. Moreover, the consideration of relativistic effects is also found to improve in the energetics of the reaction significantly. The improvement observed in the ORR activity illustrates that there could be a stronger dependence on modelled catalyst activity on relativistic effects while going to the regime of other heavy metals exhibiting considerable spin orbit coupling. Combining dispersion and relativistic effects, a maximum enhancement of ORR limiting potential by ≈ 0.2V is observed which implies the importance of consideration of these effects. Also apart from the surface studies, we believe that the ORR activity of nanoparticle based catalysts may also benefit by scrutinizing dispersion and relativistic effects.

## ASSOCIATED CONTENT

### Supporting Information

The supporting information is available free of charge on the ACS Publications website.

Table of comparison of binding energy values calculated with and without SOC and different DFT methods, free energy diagrams of ORR elementary steps with and without SOC and different DFT methods at equilibrium potential of 1.23 V.

**Conflicts of interest**

The authors declare no conflicts of interest

**Acknowledgements:**

We thank IIT Indore for providing the lab/computational facilities and DST SERB (EMR/2015/002057) and CSIR (01(2886)/17/EMR-II) projects for funding. ASN thank Ministry of Human Resources and Development, India for research fellowship.

**Table of Content:**

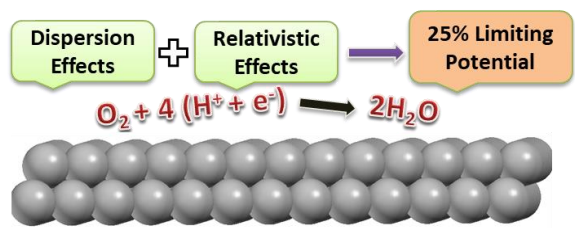